\documentclass[aps, preprint, showpacs, showkeys, amsmath, floatfix]{revtex4}
\usepackage[dvipdfmx]{epsfig}

\begin{document}

\title{Stochastic approach to generalized Schr{\"o}dinger equation:\\
        A method of eigenfunction expansion}

\date{\today}

\author{Satoshi Tsuchida}
\email[E-mail: ]{stsuchida88@gmail.com}
\affiliation{Department of Physics, Ritsumeikan University-BKC, Noji, Kusatsu City, 525--8577, Shiga, Japan\\}
\author{Hiroshi Kuratsuji}
\affiliation{Department of Physics, Ritsumeikan University-BKC, Noji, Kusatsu City, 525--8577, Shiga, Japan\\}

\begin{abstract}

Using a method of eigenfunction expansion, a stochastic equation is developed for the
generalized Schr{\"o}dinger equation with random fluctuations.
The wave field $ {\psi} $ is expanded in terms of eigenfunctions:
$ {\psi} = \sum_{n} a_{n} (t) {\phi}_{n} (x) $, with $ {\phi}_{n} $ being the eigenfunction
that satisfies the eigenvalue equation $ H_{0} {\phi}_{n} = {\lambda}_{n} {\phi}_{n} $, where
$ H_{0} $ is the reference ``Hamiltonian'' conventionally called ``unperturbed'' Hamiltonian.
The Langevin equation is derived for the expansion coefficient $ a_{n} (t) $,
and it is converted to the Fokker--Planck (FP) equation for a set $ \{ a_{n} \} $
under the assumption of the Gaussian white noise for the fluctuation.
This procedure is carried out by a functional integral,
in which the functional Jacobian plays a crucial role for determining the form of the FP equation.
The analyses are given for the FP equation by adopting several approximate schemes.

\end{abstract}


\maketitle

\section{Introduction}

The phenomenon of random fluctuations has been one of the central subjects in statistical physics,
with topics covering a wide class of physical systems ranging from molecular level to cosmological phenomena
\cite{Chandra,Uhlen,Kubo,Kadanoff}.
Brownian motion, which is one such system, is formulated by the Langevin equation,
which is an equation of motion that is modified by adding random force.
The Fokker--Planck (FP) equation, which is written for the statistical distribution function for random variables,
is an alternative form to describe random systems \cite{Chandra,Uhlen,Kubo,Kadanoff}.

The purpose of this article is to study the stochastic analyses of the generalized Schr{\"o}dinger equation in the presence of random fluctuations.
The work was inspired by previous studies of the Landau--Ginzburg (LG) equation, including
the effect of thermal agitation occurring in the superconductivity near the transition temperature (see, for example, \cite{Schmid,Larkin}).
We consider this specific problem in the framework of the wider discipline in such a way
that it can be extended to the general class of the Schr{\"o}dinger--type wave equation.
Indeed, among general wave equations, we mention wave propagation in the media
in the presence of various sorts of fluctuations caused by external agents (see e.g. \cite{Segev,Ishimaru}).
Our starting equation is a Schr{\"o}dinger--type equation with additional random force,
which is regarded as a Langevin--type equation.
Apart from the random fluctuation driven by the external agents,
the generalized Schr{\"o}dinger--type equation has been studied in the framework of the random potential \cite{Fibich,Zittarz}.
The study presented in Refs. \cite{Fibich,Zittarz} deals with the randomness that is intrinsic to the media itself, and it
belongs to a different category from the one that will be treated in what follows. 

Although it is easy to write the Langevin equation formally for the wave field $ \psi(x,t) $ itself,
the concrete calculation is rather cumbersome, when one considers applications to actual problems.
Thus, it is desirable to find a tractable way to apply the calculation.
We propose here a method of eigenfunction expansion
that is based on the use of an orthonormal set of functions that are the eigenfunctions of a {\it reference} Hamiltonian,
$ H_{0} $, which is taken as an unperturbed part of the Hamiltonian appearing in the Schr{\"o}dinger--type equation.
The method was previously developed in connection with the semi-classical analyses of a path integral \cite{Levit,Schulman},
so its application to our stochastic problem is very natural.
Therefore, the wave field is expanded by this set of eigenfunctions, and the Langevin--type equation for the
original wave field yields the Langevin--type equation for the expansion coefficients (see the next section).
This Langevin equation is converted to the FP equation by using the functional integral formalism
on the basis of Gaussian white noise for the fluctuation.
The essential point is that a functional Jacobian appearing in the functional integral plays
a key role in deriving the FP equation.

Here, we remark on previous studies of the stochastic approach to the Schr{\"o}dinger equation.
In Refs. \cite{Someda, Biele},
the stochastic differential equation is employed to describe the various sorts of
random fluctuations inherent in quantum waves.
In particular, in Ref. \cite{Someda}, this equation was used to derive the FP equation.
The method developed therein may not always provide relevant tools for the wide class of problems described by the generalized Schr{\"o}dinger equation.
However, the present approach, in which the eigenfunction expansion is used, is expected to provide a very simple and general
technique for studying the stochastic characteristics of the generalized Schr{\"o}dinger equation.

The paper is organized as follows:
In the next section, the Langevin equation is derived for the generalized Schr{\"o}dinger equation.
In Sec. \ref{sec:functionalint}, the Langevin equation is converted to the functional integral on the basis of Gaussian white noise.
In Sec. \ref{sec:fokkerplanck}, the FP equation is derived from the functional integral by adopting the trick of imaginary time.
This is based on the well--known procedure for deriving the Schr{\"o}dinger equation from a path integral.
In Sec. \ref{sec:analysesfp}, the analyses of the FP equation are presented for the case 
in which the expansion coefficients are treated as independent of each other.
In Sec. \ref{subsec:strongcouple},
we adopt the strong--coupling approximation to derive the expression for the ``first--excited'' state by using the variational method.
In Sec. \ref{subsec:smalldiffuse}, we present the transition amplitude for the small diffusion limit, which corresponds to the semi-classical approximation.

\section{Langevin equation for the Schr{\"o}dinger--type equation}
\label{sec:langevinsch}

Our starting point is the Langevin type generalized Schr{\"o}dinger equation for the wave field $ {\psi} (x,t) $ \cite{Schmid, Larkin}:
\begin{eqnarray}
  \label{eq:langevinpsi}
  {\gamma} \frac{ {\partial} {\psi} (x,t) }{ {\partial} t } = H {\psi} (x,t) + {\eta} (x,t),
\end{eqnarray}
where $ H $ is an operator corresponding to ``Hamiltonian''.
The coefficient $ {\gamma} $ denotes the complex diffusion constant,
and $ {\eta} (x,t) $ represents fluctuation coming from thermal
and other effects, which obeys the law of white noise,
\begin{eqnarray}
  \label{eq:whitenoise}
  \langle {\eta}^{*} (x,t) {\eta} (x',t') \rangle = {h} {\delta} (x-x') {\delta} (t-t'),
\end{eqnarray}
where $ h $ is a diffusion constant for the random process.

In the following discussion, $ H $ is assumed to be given by the sum: $ H = H_{0} + V = ( {\nabla}^{2} + U ) + V $.
$ H_{0} $ is conventionally called an unperturbative term described by the potential $ U $, whereas
$ V $ represents a perturbation, which allows time dependence in general.
$ H_{0} $ and $ V $ are assumed to be Hermitian.

Now let us consider the eigenfunction expansions for $ {\psi} $ and $ {\eta} $,
\begin{eqnarray}
  \label{eq:eigenpsi}
  {\psi} (x,t) = \sum_{n} a_{n}(t) {\phi}_{n}(x) \ \ , \ \ {\eta} (x,t) = \sum_{n} {\eta}_{n}(t) {\phi}_{n}(x),
\end{eqnarray}
where $ a_{n} (t) $ and $ {\eta}_{n} (t) $ are complex functions of $ t $, and ${\phi}_{n} (x)$ is defined as an eigenfunction
$ H_{0} {\phi}_{n} (x) = {\lambda}_{n} {\phi}_{n} (x) $, where we assume
the eigenvalue $ {\lambda}_{n} $ is positive for the convenience of later arguments.
By using the orthogonality relation
$ \int {\phi}_{n} (x) {\phi}_{m}^{*} (x) dx = {\delta}_{nm} $,
we can get a Langevin equation for $ a_{n} (t) $,
\begin{eqnarray}
  \label{eq:langevin}
  {\gamma} \frac{{\partial}a_{n}(t)}{{\partial}t} = \frac{ {\partial} E }{ {\partial} a_{n}^{*} } + {\eta}_{n}(t),
\end{eqnarray}
together with the complex conjugate.
Here we introduce the {\it energy function} $ E $, which consists of the unperturbed and the perturbation terms:
$ E = E_{0} + E_{v} $, with
\begin{eqnarray}
  \label{eq:energy}
  E_{0}  & = & \sum_{n} \lambda_{n} a_{n}^{*} a_{n},  \nonumber \\
  E_{v} & = & \sum_{n,m} \langle n \vert \hat{V} \vert m \rangle a_{n}^{*} a_{m},
\end{eqnarray}
where $ V_{nm} = V^{*}_{mn} $ (that is, $ V $ denotes the Hermitian),
and $ \vert n \rangle $ and $ \vert m \rangle $ are the eigenstates for mode $ n $ and $ m $, respectively.
The complex variable $ a_{n} $ can be expressed in terms of real and imaginary parts as $ a_{n} = X_{n} + i Y_{n} $,
which is used for the concrete description of the FP equation.

Here we remark on the complex coefficient $ {\gamma} $ in (\ref{eq:langevin}):
For this purpose, we rewrite Eq. (\ref{eq:langevin}) in terms of real variables $ (X, Y) $:
\begin{eqnarray}
  \label{eq:langevinrealimage}
  {\alpha} \frac{ dX_{n} }{ dt } + {\kappa} \frac{ dY_{n} }{ dt } = - \frac{1}{2} \frac{ {\partial} E }{ {\partial} X_{n} } + {\rm Re}(\eta),  \nonumber \\
  {\alpha} \frac{ dY_{n} }{ dt } - {\kappa} \frac{ dX_{n} }{ dt } = - \frac{1}{2} \frac{ {\partial} E }{ {\partial} Y_{n} } + {\rm Im}(\eta).
\end{eqnarray}
Here $ {\alpha} $ and $ {\kappa} $ are defined by $ {\gamma} = - {\alpha} + i {\kappa} $.
If we set $ {\kappa} = 0 $, this equation describes the dissipation, hence $ {\alpha} $ should be positive.
On the other hand, for the opposite case in which $ {\alpha} = 0 $,
one can see that the equation of motion leads to the non-dissipative motion, since the energy function $ E $ is conserved.

\section{Functional Integral}
\label{sec:functionalint}

According to the assumption of white noise, $ {\eta} ( x, t ) $ obeys the Gaussian distribution;
\begin{eqnarray}
  \label{eq:whitenoiseeta}
  P \left[ {\eta}(x,t) \right] = N {\rm{exp}} \left[ - \frac{1}{2h} \int {\eta}^{*}(x,t) {\eta}(x,t) dx dt \right],
\end{eqnarray}
where $ h $ is the diffusion constant corresponding to the Planck constant.
By considering the Langevin equation and taking account of the
boundary condition: $ {\psi}(x,0) \equiv \psi(x) $, $ {\psi}(x,T) \equiv \psi'(x) $,
the transition probability from $ {\psi} (x) $ to $ {\psi}' (x) $ is written by the path integral,
\begin{eqnarray}
  \label{eq:transitionprob}
  K( {\psi}'(x), T \vert {\psi}(x), 0 ) \sim \int_{\psi(x)}^{\psi' (x)} {\rm{exp}} \left[ - \frac{1}{2 h} \int_{0}^{T} {\eta}^{*} (x,t) {\eta}(x,t) dx dt \right] \prod_{x,t} {\cal{D}} {\eta(x,t)}^{*} {\cal{D}} {\eta(x,t)},
\end{eqnarray}
which turns out to be represented in terms of the amplitude $ \{ a_{n} \} $,
\begin{eqnarray}
  \label{eq:functionalintan}
  K( \{ a_{n}' \}, T \vert \{ a_{n} \}, 0) = \int_{ \{ a_{n} \} }^{ \{ a_{n}' \} } {\rm{exp}} \left[ - \frac{1}{2 h} \int_{0}^{T} \sum_{n} {\eta}^{*}_{n} (t) {\eta}_{n} (t) dt \right] 
  \prod_{n,t}  {\cal{D}} { {\eta}_{n} (t) }^{*} {\cal{D}} { {\eta}_{n} (t) },
\end{eqnarray}
where we use the eigenfunction expansion defined by Eq.(\ref{eq:eigenpsi}).
The functional integral (\ref{eq:functionalintan}) is rewritten in the following steps:
First, by inserting the expression for the identity of the $ \delta $--functional integral \cite{Zinn,Ramond}:
\begin{eqnarray}
  \label{eq:deltafunctional}
  \int \prod_{n} \prod_{t} {\delta} \left[ F_{n}(t) - {\eta}_{n}(t) \right] {\cal{D}} F_{n}(t) = 1 \ \ \ \left( F_{n}(t) = {\gamma} \frac{{\rm{d}}a_{n}}{{\rm{d}}t} - \frac{ {\partial} E }{ {\partial} a_{n}^{*} } \right)
\end{eqnarray}
together with use of the well--known relation for $ {\delta} $ function:
$ \delta[ f(x) ] \propto \int_{- \infty}^{\infty} {\rm{exp}} \left[ i {\rho} f(x) \right] d {\rho} $,
and next by integrating over $ {\eta}_{n} $ ( $ \eta_{n}^{*} $ ) as well as $ {\rho}_{n} $ ( $ {\rho}_{n}^{*} $ ), we obtain
\begin{eqnarray}
  \label{eq:pathintegralforfn}
   K \sim  C  \int {\rm{exp}} \left[ - \frac{1}{2h} \int_{0}^{T} \sum_{n} F_{n}^{*} F_{n} dt \right] {\cal{D}} F_{n}^{*} {\cal{D}} F_{n},
\end{eqnarray}
where $ C $ is a factor that does not depend on $ F_{n} $.
By the definition of $ F_{n} $ and by using a functional determinant,
the path integral is given as
\begin{eqnarray}
  \label{eq:resultingpath}
  K \sim  C  \int {\rm{exp}} \left[ - \frac{1}{2h} \int_{0}^{T} \sum_{n} \left( {\gamma}^{*} \frac{da_{n}^{*}}{dt} - \frac{ {\partial} E }{ {\partial} a_{n} } \right) \left( {\gamma} \frac{da_{n}}{dt} - \frac{ {\partial} E }{ {\partial} a_{n}^{*} } \right) dt \right] \nonumber \\
  \times {\rm{det}} \left[ \frac{{\delta}F_{n}^{*}}{{\delta}a_{n}^{*}} \right] {\rm{det}} \left[ \frac{{\delta}F_{n}}{{\delta}a_{n}} \right] {\cal{D}}a_{n}^{*} {\cal{D}}a_{n}.
\end{eqnarray}
The functional determinant is calculated and written by using real variables $ (X, Y) $
\begin{eqnarray}
  \label{eq:functionaljacobi}
  {\rm{det}} \left[ \frac{{\delta}F_{n}}{{\delta}a_{n}} \right] = {\rm{exp}} \left[ - \frac{1}{ 8 \gamma} \int_{0}^{T} \left( \frac{ {\partial}^{2} E }{ {\partial} X_{n}^{2} } + \frac{ {\partial}^{2} E }{ {\partial} Y_{n}^{2} } \right) dt \right]
\end{eqnarray}
This factor plays a crucial role in fixing the correct form of the FP equation.
The process to derive it will be given in the Appendix \ref{append:functionaljacobi}.

From the above form of path integral, we see that an action functional is modified as 
\begin{eqnarray}
  \label{eq:action}
  S &=& \sum_{n} \int_{0}^{T} \frac{1}{2} \Big{|} {\gamma} \frac{da_{n}}{dt} - \frac{ {\partial} E }{ {\partial} a_{n}^{*}} \Big{|}^{2} dt + \frac{h}{8} \left( \frac{1}{{\gamma}^{*}} + \frac{1}{\gamma} \right) \sum_{n} \int_{0}^{T} \left( \frac{ {\partial}^{2} E }{ {\partial} X_{n}^{2} } + \frac{ {\partial}^{2} E }{ {\partial} Y_{n}^{2} } \right) dt  \nonumber \\
        &=& \sum_{n} \left( \int_{0}^{T} {\tilde{{\cal{L}}}}_{n} dt + h \int_{0}^{T} M dt \right),
\end{eqnarray}
where
\begin{eqnarray}
  \label{eq:jacobiresult}
  M = - \frac{\alpha}{ 4 |{\gamma}|^{2} } \left( \frac{ {\partial}^{2} E }{ {\partial} X_{n}^{2} } + \frac{ {\partial}^{2} E }{ {\partial} Y_{n}^{2} } \right).
\label{eq:M}
\end{eqnarray}

Now we rewrite the Lagrangian ${\tilde{\cal{L}}}_{n}$ in Eq.(\ref{eq:action}) by using real coordinates, $ (X,Y) $,  explicitly,
and further by noting ${\gamma} = - {\alpha} + i {\kappa}$. Then we obtain
\begin{eqnarray}
  \label{eq:lagrangiantilde}
  {\tilde{\cal{L}}}_{n} &=& \frac{1}{2} \Biggl\{ |{\gamma}|^{2} \left( {\dot{X}_{n}}^{2} + {\dot{Y}_{n}}^{2} \right) + {\alpha} \left( {\dot{X}_{n}} \frac{ {\partial} E }{ {\partial} X_{n}} + {\dot{Y}_{n}} \frac{ {\partial} E }{ {\partial} Y_{n}} \right) \nonumber \\
    &-& {\kappa} \left( {\dot{X}_{n}} \frac{ {\partial} E }{ {\partial} Y_{n}} - {\dot{Y}_{n}} \frac{ {\partial} E }{ {\partial} X_{n}} \right) + \frac{1}{4} \left[ \left( \frac{ {\partial} E }{ {\partial} X_{n} } \right)^{2} + \left( \frac{ {\partial} E }{ {\partial} Y_{n}} \right)^{2} \right] \Biggr\}.
\end{eqnarray}

Here in order to make a connection with quantum mechanics,
we introduce imaginary time $ {\tau} ( = - it ) $,
and we set the Lagrangian as $ {\cal{L}}_{n} ( = - \tilde{{\cal{L}}}_{n}) $.
By using the conjugate momentum $ {\bf{p}}_{n} = \frac{{\partial}{\cal{L}}_{n}}{{\partial}{\dot{\bf{X}}}_{n}} $
(${\dot{{\bf{X}}}}_{n} = \frac{d{\bf{X}}_{n}}{d{\tau}}$),
the Hamiltonian corresponding to $ {\cal{L}}_{n} $ becomes
\begin{eqnarray}
  \label{eq:hamiltonianpn}
  {\cal{H}}_{n} &=& \frac{1}{2 |{\gamma}|^{2}} \left[ \left( p_{nX} - i {\alpha} \frac{ {\partial} E }{ {\partial} X_{n} } + i {\kappa} \frac{ {\partial} E }{ {\partial} Y_{n}} \right)^{2} + \left( p_{nY} - i {\alpha} \frac{ {\partial} E }{ {\partial} Y_{n} } - i {\kappa} \frac{ {\partial} E }{ {\partial} X_{n} } \right)^{2} \right]    \nonumber \\
  &+& \frac{1}{8 |{\gamma}|^{2} } \left[ \left( \frac{ {\partial} E }{ {\partial} X_{n} } \right)^{2} + \left( \frac{ {\partial} E }{ {\partial} Y_{n} } \right)^{2} \right].
\end{eqnarray}
Thus, the total Hamiltonian is given by
$ {\cal{H}} = \sum_{n} {\cal{H}}_{n}. $

\section{The Fokker--Planck equation}
\label{sec:fokkerplanck}

We now derive the FP equation via the ``imaginary time'' Schr{\"o}dinger equation
by starting with a quantized version of the Hamiltonian (\ref{eq:hamiltonianpn}).
First we note that the corresponding ``wave function'' $ {\Psi} ( \{ {\bf{X}}_{n} \}, {\tau} ) $ is related
with the probability distribution function $ P ( \{ {\bf{X}}_{n} \}, t ) $ as follows:
\begin{eqnarray}
  \label{eq:wavepro}
  {\Psi} ( \{ {\bf{X}}_{n} \}, \tau ) = P( \{ {\bf{X}}_{n} \}, t ).
\end{eqnarray}
As is well known, the Schr{\"o}dinger equation is given by integral equation
\begin{eqnarray}
  \label{eq:integraleq}
  {\psi} ( \{ {\bf{X}}_{n} \}, {\tau} + {\epsilon} ) = \int K( \{ {\bf{X}}_{n} \}, {\tau} + {\epsilon} \ \vert \ \{ {\bf{X}}_{n}' \}, {\tau} ) {\psi} ( \{ {\bf{X}}_{n}' \} , {\tau} ) \prod_{n} d{\bf{X}}_{n}', \nonumber
\end{eqnarray}
where the propagator $ K $ is constructed from the action functional given above.
Thus, according to the well--known procedure (see, \cite{Hibbs}), the Schr{\"o}dinger equation for the
wave function $ {\Psi} ( \{ {\bf{X}}_{n} \}, {\tau} ) $ turns out to be
\begin{eqnarray}
  \label{eq:schrotypeeq}
  ih \frac{{\partial}{\Psi}}{{\partial}{\tau}} &=& \sum_{n} \frac{1}{2|{\gamma}|^{2}} \left[ \left( p_{nX} - i \frac{\alpha}{2} \frac{ {\partial} E }{ {\partial} X_{n} } 
      + i \frac{\kappa}{2} \frac{ {\partial} E }{ {\partial} Y_{n} } \right)^{2} + \left( p_{nY} - i \frac{\alpha}{2} \frac{ {\partial} E }{ {\partial} Y_{n} } - i \frac{\kappa}{2} \frac{ {\partial} E }{ {\partial} X_{n} } \right)^{2} \right] {\Psi}  \nonumber \\
  &+& \sum_{n} W {\Psi},
\end{eqnarray}
where
\begin{eqnarray}
  \label{eq:pandvrelation}
  p_{nX} = -ih \frac{{\partial}}{{\partial}X_{n}} \ \ , \ \ p_{nY} = -ih \frac{{\partial}}{{\partial}Y_{n}},  \nonumber \\
  W = \frac{1}{8} \left[ \left( \frac{ {\partial} E }{ {\partial} X_{n} } \right)^{2} + \left( \frac{ {\partial} E }{ {\partial} Y_{n} } \right)^{2} \right] + M h.
\end{eqnarray}
Then, by using Eq.(\ref{eq:pandvrelation}), and replacing the imaginary time $ \tau $ with the original real time $ t $,
namely, $ {\tau} \to - it $, we obtain 
\begin{eqnarray}
  \label{eq:schroe}
  \frac{{\partial}{\Psi}}{{\partial}t} = \sum_{n} \Biggl\{ \frac{h}{2|{\gamma}|^{2}} \left[ \left( \frac{{\partial}}{{\partial}X_{n}} \right)^{2} + \left( \frac{{\partial}}{{\partial}Y_{n}} \right)^{2} \right] {\Psi}  \nonumber \\
+ \frac{1}{2 |{\gamma}|^{2}} \left( {\alpha} \left[ \frac{ {\partial} E }{ {\partial} X_{n} } \frac{{\partial}{\Psi}}{{\partial}X_{n}} + \frac{ {\partial} E }{ {\partial} Y_{n} } \frac{{\partial}{\Psi}}{{\partial}Y_{n}} \right] + {\kappa} \left[ \frac{ {\partial} E }{ {\partial} X_{n} } \frac{{\partial}{\Psi}}{{\partial}Y_{n}} - \frac{ {\partial} E }{ {\partial} Y_{n} } \frac{{\partial}{\Psi}}{{\partial}X_{n}} \right] \right) \\
+ \left[ \frac{\alpha}{ 4 |{\gamma}|^{2}} \left( \frac{ {\partial}^{2} E }{ {\partial} X_{n}^{2} } + \frac{ {\partial}^{2} E }{ {\partial} Y_{n}^{2} }  \right) - M \right] {\Psi} \Biggr\}. \nonumber
\end{eqnarray}
Using (\ref{eq:M}), the last term becomes
$ \frac{\alpha}{ 2 |{\gamma}|^{2}} \left( \frac{ {\partial}^{2} E }{ {\partial} X_{n}^{2} } + \frac{ {\partial}^{2} E }{ {\partial} Y_{n}^{2} }\right)  $. 
Hence the second and the last terms can be combined in a single form by noting the relation
$ \nabla {\bf{A}} \cdot f + {\bf{A}} \cdot \nabla f = \nabla \cdot ( {\bf{A}} f ) $.
Further, by getting the wave function $ {\Psi} $ back to the original probability distribution $ P $, we finally obtain the FP equation: 
\begin{eqnarray}
  \label{eq:FP}
  \frac{{\partial}P}{{\partial}t} = \sum_{n} \Big[ \frac{h}{2|{\gamma}|^{2}} \left( \frac{{\partial}}{{\partial}X_{n}} \right)^{2} P + \frac{1}{ 2 |{\gamma}|^{2}} \frac{\partial}{{\partial}X_{n}} \left( {\alpha} \frac{ {\partial} E}{ {\partial} X_{n} } P - {\kappa} \frac{ {\partial} E }{ {\partial} Y_{n} } P \right)  \nonumber \\
   + \frac{h}{2|{\gamma}|^{2}} \left( \frac{{\partial}}{{\partial}Y_{n}} \right)^{2} P + \frac{1}{ 2 |{\gamma}|^{2}} \frac{\partial}{{\partial}Y_{n}} \left( {\alpha} \frac{ {\partial} E }{ {\partial} Y_{n} } P + {\kappa} \frac{ {\partial} E }{ {\partial} X_{n} } P \right) \Big].
\end{eqnarray}

Here we discuss some specific features of this equation:

(i): Equation (\ref{eq:FP}) can be represented in a continuity equation of the probability:
\begin{eqnarray}
  \label{eq:continuity}
  \frac{{\partial}P}{{\partial}t} + \nabla \cdot J = 0,
\end{eqnarray}
where the ``current'' is defined as $ J = \sum_{n} J_{n} $,
\begin{eqnarray}
  \label{eq:current}
  J_n = - \frac{1}{2 |{\gamma}|^{2} } \left( h {\nabla}_{n} P + {\bf{A}}_{n} P \right) \ \ , \ \ 
  {\bf{A}}_n  = {\alpha} \frac{\partial E}{\partial {\bf X}_n} - {\kappa} {\bf k} \times \frac{\partial E}{\partial {\bf X}_n},
\end{eqnarray}
with $ {\bf{k}} $ being the unit vector that is perpendicular to the $ ( X, Y ) $ plane.
Indeed, it has been known that the FP equation can be derived by
following an analogy with the continuity equation for the probability flow in an intuitive way \cite{Brown}.

(ii): As a special case, we consider stationary distribution, namely
the case of $ \frac{ {\partial} P }{ {\partial} t } = 0 $, for which we propose
\begin{eqnarray}
  \label{eq:ansatz}
  P = e^{ - {\beta} E },
\end{eqnarray}
where $ \beta = 1 / kT $ denotes the inverse temperature. Substituting this form, we have 
\begin{eqnarray}
  \label{eq:FDT}
  ( {\alpha} - h {\beta} ) \sum_{n} \left\{ \frac{ {\partial}^{2} E }{ {\partial} X_{n}^{2} } + \frac{ {\partial}^{2} E }{ {\partial} Y_{n}^{2} }
   - {\beta} \left( \left( \frac{ {\partial} E }{ {\partial} X_{n} } \right)^{2} + \left( \frac{ {\partial} E }{ {\partial} Y_{n} } \right)^{2} \right) \right\} = 0.
\end{eqnarray}
For this relation to be satisfied for arbitrary function $ E $, the following relation should be held:
\begin{eqnarray}
  \label{eq:equilibrium}
  {\beta} = \frac{\alpha}{h}.
\end{eqnarray}
This is merely the fluctuation dissipation theorem,
which establishes the relation among three parameters $ {\alpha} $, $ {\beta} $, and $ h $,
representing the dissipation, inverse temperature, and diffusion, respectively.
Note that the effects of $ {\kappa} $ disappear in the above relation (\ref{eq:FDT}),
and non-zero $ {\alpha} $ plays an essential role for the existence of the equilibrium state.
If $ {\alpha} = 0 $, the equilibrium state does not exist,
because the effective temperature becomes infinite.

{\it{Calculation of the partition function}}:
The equilibrium distribution is used to evaluate the partition function, which leads to various sorts of thermodynamical quantities
\cite{Feynman}: 
As a particular case, we consider the case in which the energy function is given by the unperturbed one, $ E_0 = \sum_n \lambda_n (X_n^2 + Y_n^2) $,
for which we have 
\begin{eqnarray}
  \label{eq:partition}
  Z = \int \exp[-\beta \sum_n \lambda_n a_n^{*}a_n ]\prod_n da_n^{*}da_n
  = \prod_{n} \left[ \frac{ 2 {\pi} }{ {\beta} {\lambda}_{n} } \right].
\end{eqnarray}
This is just the same form as the partition function used in the superconductivity near the phase transition \cite{Schmid,Langer},
which is described by the time--independent LG free energy. That is, the partition function is given by
$ Z = \int \exp \left[ - {\beta} F \left[ \psi,\psi^{*} \right] \right] \prod d\psi^{*}d\psi $ with $ F[\psi,\psi^{*} ] = \int \psi^{*}H \psi dx $.
$ H $ is given as the Hamiltonian that depends on the electromagnetic field. 
In this way, the present approach includes the stationary problem for the LG theory as a special case.  

(iii): The FP equation describes the coupling among the modes $ a_{i} \ (i = 1, \cdots, \infty $).
This is governed by the perturbation term $ E_v $ in the energy function (\ref{eq:energy}), which is written in terms of the real variables $ (X_n,Y_n) $: 
\begin{eqnarray}
  \label{eq:evrealima}
  E_{v} = \sum_{n,m} B_{nm} ( X_{n} X_{m} + Y_{n} Y_{m} ) + C_{nm} ( X_{n} Y_{m} - X_{m} Y_{n} ),
\end{eqnarray}
where $ B $, $ C $ are the real and imaginary parts of the hermitian matrix $ V_{nm} $: ( $ V_{nm} = B_{nm} + i C_{nm} $);
hence these become the symmetric and skew-symmetric matrices respectively, namely, $ B_{nm} = B_{mn} $, $ C_{nm} = - C_{mn} $.

In particular, the FP equation described by the unperturbed energy $ E_{0} $ yields
\begin{eqnarray}
  \label{eq:unperturb}
  \frac{{\partial}P}{{\partial}t} = \sum_{n} \Biggl[ \frac{h}{2|{\gamma}|^{2}} \left( \frac{{\partial}}{{\partial}X_{n}} \right)^{2} P 
    + \frac{{\lambda}_{n}}{|{\gamma}|^{2}} \frac{\partial}{{\partial}X_{n}} \left( {\alpha} X_{n} P - {\kappa} Y_{n} P \right) \nonumber \\
    + \frac{h}{2|{\gamma}|^{2}} \left( \frac{{\partial}}{{\partial}Y_{n}} \right)^{2} P 
    + \frac{{\lambda}_{n}}{|{\gamma}|^{2}} \frac{\partial}{{\partial}Y_{n}} \left( {\alpha} Y_{n} P + {\kappa} X_{n} P \right) \Biggr].
\end{eqnarray}
Furthermore, the term coming from $ E_{v} $ gives rise to the modification to the above; that is,
\begin{eqnarray}
  \label{eq:perteffect}
  \sum_{nm} \frac{1}{ |{\gamma}|^{2} } \frac{\partial}{\partial X_{n}} \Big[ {\alpha} \left( B_{lm} X_{m} + C_{lm} Y_{m} \right) P
                                                                           - {\kappa} \left( B_{lm} Y_{m} - C_{lm} X_{m} \right) P \Big] \nonumber \\
  + \sum_{nm} \frac{1}{ |{\gamma}|^{2} } \frac{\partial}{\partial Y_{n}} \Big[ {\alpha} \left( B_{lm} Y_{m} - C_{lm} X_{m} \right) P
                                                                           + {\kappa} \left( B_{lm} X_{m} - C_{lm} Y_{m} \right) P \Big].
\end{eqnarray}
This term plays the role of the perturbation to the unperturbed equation (\ref{eq:unperturb}).
Also the coefficients $ B_{nm} $, $ C_{nm} $ have time-dependence in general, so
we have a time-dependent perturbation theory for the FP equation
that may be expected to bring about specific physical consequences.

\section{Analyses of the stochastic equation}
\label{sec:analysesfp}

The FP equation that was derived above is described by the energy function $ E $,
which includes the coupling among various modes.
In the following discussion, we restrict the argument to the specific case of the independent mode
$ E_{0}  = \sum_{n} {\lambda}_{n} a_{n}^{*} a_{n} $.
For this case, the stochastic equation is given by (\ref{eq:unperturb}),
and the analyses of this will be carried out in a simple manner, because the modes are decoupled from each other.
That is, the distribution function is written as a product form:
\begin{eqnarray}
  \label{eq:distrifun}
  P \left( \{ {\bf{X}} \}, t \right) = \prod_{n} P_{n} ( X_{n}, Y_{n}, t ), \nonumber
\end{eqnarray}
where $ P_{n} ( X_{n}, Y_{n}, t ) $ means the distribution for the $n$-th mode,
and the term including $ {\kappa} $ is omitted because it does not contribute to the equilibrium state.
In what follows, we consider the $n$-th mode only, so the index $n$ is suppressed.
Then, we set the variable separation as
\begin{eqnarray}
  \label{eq:varisep}
  P(X,Y,t) = \exp [ - {\epsilon} t ] f(X,Y).  \nonumber
\end{eqnarray}
Thus, we get the eigenvalue problem as follows:
\begin{eqnarray}
  \label{eq:eigenpro}
  \left( \frac{ {\partial}^2 }{ {\partial} X^2 } + \frac{ {\partial}^2 }{ {\partial} Y^2} \right) f + \frac{ 2 \lambda \alpha}{h} \left( X \frac{ {\partial} f }{ {\partial} X } + Y \frac{ {\partial} f }{ {\partial} Y } \right)
  = - \frac{ 2 {\alpha} }{ h } \left( {\alpha} {\epsilon} + 2 {\lambda} \right) f.
\end{eqnarray}
Here we use the polar coordinates: $ X = R \cos \theta $, $ Y = R \sin \theta $,
and we consider the case in which the eigenfunction $ f(X,Y) $ does not depend on the angular variable $ {\theta} $.
Therefore, the eigenvalue problem becomes
\begin{eqnarray}
  \label{eq:eigenmu}
  \frac{1}{R} \frac{d}{dR}
  \left[ R \left( \frac{df}{dR} + \frac{\mu}{2} \frac{dU}{dR} f \right) \right] = - \frac{ 2 {\alpha}^{2} }{ h } {\epsilon} f, 
\end{eqnarray}
where $ {\mu} = \frac{ 2 {\lambda} {\alpha} }{h} = 2 {\lambda} {\beta} $.
Here we use the notation $ U(R) = X^2 + Y^2 = R^2 $, then Hamiltonian is written as $ H = {\lambda} U(R) $.

\subsection{Strong coupling approximation}
\label{subsec:strongcouple}

The eigenvalue problem (\ref{eq:eigenmu}) looks simple,
but this may not be represented by special functions that have been used so far.
If we note that Eq.(\ref{eq:eigenmu}) includes the parameter $ \mu $, which can be regarded as a perturbation parameter,
one may think of carrying out the perturbation scheme by expansion with respect to the small parameter $ \mu $ ($ \vert \mu \vert \ll 1 $).
That is, one starts with the equation for $ \mu = 0 $ as an unperturbed solution, which is given by the Bessel function.
However this procedure may not be relevant, since the case
$ \mu = 0 $ corresponds to $ \alpha = 0 $, which does not represent thermodynamic equilibrium as it is pointed out above.
From this inspection, it is suitable to consider the case in which $ |{\mu}| $ is not small,
$ \vert \mu \vert \simeq 1 $, which we call the ``strong coupling approximation''.
The following procedure is similar to that used in the stochastic approach of ferromagnetic 
particles \cite{Brown}, although the problem under consideration belongs to a completely different discipline. 

Then, we immediately get the solution for the zero eigenvalue ($ \epsilon = 0 $):
\begin{eqnarray}
  \label{eq:f0}
  f_0(R) = \exp \left[ - \frac{\mu}{2} U(R) \right].
\end{eqnarray}
This corresponds to the Boltzmann distribution, which is identified with the ``ground state''.
As a sequel of the ground state we assume ``excited state'' as
\begin{eqnarray}
  \label{eq:fr}
  f(R) = \exp \left[ - \frac{\mu}{2} U(R) \right] g(R).
\end{eqnarray}
The function $ g (R) $ satisfies
\begin{eqnarray}
  \label{eq:eigenvaluepro}
  \frac{d}{dR}\left( R \exp \left[ - \frac{\mu}{2} U \right] \frac{dg}{dR} \right) = \frac{ 2 {\alpha}^{2} }{h} {\epsilon} R \exp \left[ - \frac{\mu}{2} U \right] g.
\end{eqnarray}
In particular, for $ \epsilon = 0 $, we can set $ \frac{dg}{dR} $ to be zero,
which gives $ g_{0} (R) = k $ (constant).

The eigenvalue equation (\ref{eq:eigenvaluepro}) is equivalent to the following variational problem:
That is, the functional given by
\begin{eqnarray}
  \label{eq:actionfori}
  I = \int_0^{\infty}  \left( \frac{dg}{dR} \right)^{2} \exp \left[ - \frac{\mu}{2} U \right] R dR
\end{eqnarray}
should be minimized under the normalization condition
\begin{eqnarray}
  \label{eq:normalizedg}
  \int^{\infty}_0 g^2 \exp \left[ - \frac{\mu}{2} U \right] R dR = 1.
\end{eqnarray}
Furthermore, we propose that the orthogonality relation should be held
between two eigenstates $ g $ and $ {\tilde{g}} $:
\begin{eqnarray}
  \label{eq:orthogonalg}
  \int g {\tilde{g}} \exp \left[ - \frac{\mu}{2} U \right] R dR = 0.
\end{eqnarray}

Under this prescription, we consider the first--excited state as a concrete example.
To perform the variational calculation, we choose the trial function as the quadratic expression
\begin{eqnarray}
  \label{eq:trialfunc}
  g_{1} (R) = A + BR + CR^{2}.
\end{eqnarray}
The coefficients $ A $, $ B $ and $ C $ are determined by three conditions corresponding to Eqs.(\ref{eq:actionfori})--(\ref{eq:orthogonalg}),
which are written explicitly as
\begin{eqnarray}
  \label{eq:energyi}
  I  =  J_{1} B^{2} + 4 J_{2} BC + 4 J_{3} C^{2} \ ( \equiv {\epsilon}_{1} )
\end{eqnarray}
and
\begin{eqnarray}
  \label{eq:constraints}
  J_{1} A^{2} + 2 J_{2} AB + J_{3} ( 2 AC + B^{2} ) + 2 J_{4} BC + J_{5} C^{2} = 1,  \nonumber \\
  J_{1} A + J_{2} B + J_{3} C = 0.
\end{eqnarray}
respectively.
Here we set $ J_{k} = \int_{0}^{\infty} \exp \left[ - \frac{\mu}{2} R^{2} \right] R^{k} dR $.
The minimum value of $ I $ is obtained as follows:
First, by eliminating $ A $ from Eq.(\ref{eq:constraints}),
one gets the quadratic constraint with respect to $ B $, $ C $:
\begin{eqnarray}
  \label{eq:constraintbc}
  G(B,C) = \left( \frac{ J_{2}^{2} }{ J_{1} } + J_{3} \right) B^{2} + 2 J_{4} B C + \left( J_{5} - \frac{ J_{3}^{2} }{ J_{1} } \right) C^{2} - 1 = 0.
\end{eqnarray}
Next, by using this constraint and following the Lagrange multiplier method,
we have the relations
\begin{eqnarray}
  \label{eq:lagrangemulti}
  \frac{\partial}{ {\partial} {\chi}_{i} } ( I - {\lambda} G ) = 0 ~~ ( {\chi}_{i} = B, C, {\lambda}),
\end{eqnarray}
where ${\lambda}$ is a multiplier.
Equation (\ref{eq:lagrangemulti}) leads to simultaneous equations for $B$, $C$, and ${\lambda}$.
By solving this, we can obtain the value of $ {\epsilon}_{1} $.
The explicit manipulation is rather tedious and is omitted here.

By continuing the above procedure,
we can obtain a sequence of the excited states $ g_{n} (R) $.
By using this, the distribution function $ P (R,t) $ is expanded as
\begin{eqnarray}
  \label{eq:expandp}
  P(R,t) = \sum_n C_{n} (t) g_{n} (R) \exp \left[ - \frac{\mu}{2} U \right].
\end{eqnarray}
The expansion coefficient $ C_{n} (t) $ satisfies $ d C_{n} / dt = - {\epsilon}_{n} C_{n} $,
and from the initial condition $ P (R, 0) = \sum_n C_{n} (0) g_{n} (R) \exp \left[ - \frac{\mu}{2} U \right] $,
we get $ C_n (0) = \int P (R,0) g_{n} (R) \exp \left[ - \frac{\mu}{2} U \right] R dR $.
As an actual situation, it may be sufficient to keep the terms up to the first excited state,
\begin{eqnarray}
  \label{eq:excitedp}
  P(R,t) = \{ C_0(t) g_0 + C_1(t)g_1(R) \} \exp \left[ - \frac{\mu}{2} U \right].
\end{eqnarray}
The distribution of this form may be utilized to calculate the time evolution of the average 
value of physical quantities under consideration.

\subsection{Small diffusion limit}
\label{subsec:smalldiffuse}

We consider another procedure of the nonperturbational scheme
that is given as the asymptotic limit of the path integral, namely
the limit of zero diffusion, $ h \sim 0 $.
This corresponds to the semiclassical approximation in quantum mechanics.
The path integral expression for $ h \sim 0 $ becomes
\begin{eqnarray}
  \label{eq:semiclapath}
  K_{sc} = \exp \left[ - \frac{S_{sc}}{h} \right].
\end{eqnarray}
Here the classical action is written as a sum of the contributions coming from each mode $ n $; $ S_{sc} = \sum_n S_{sc}^n $.
In what follows, we take account of the mode $ n $ only, and we suppress the index $ n $.
The final result is obtained by summing over $ n $.
By using the polar coordinate, the action functional is given as
\begin{eqnarray}
  \label{eq:semiclaaction}
  S_{sc} = \int \left[ \frac{ {\alpha}^{2} }{ 2 } \left( \dot{R} - \zeta \frac{dU}{dR} \right)^{2} + \frac{ {\alpha}^{2} }{ 2 } R^{2} \dot{\theta}^{2} \right] dt,
\end{eqnarray}
where $ \zeta = \frac{\lambda}{ 2 {\alpha} } $.
The conjugate momentum for $ \theta $, $ p_{\theta} = R^{2} \dot{\theta} $, is conserved
because $ {\theta} $ is a cyclic coordinate.
Then, the Lagrangian becomes
\begin{eqnarray}
  \label{eq:lagrangianforr}
  L = \frac{ {\alpha}^{2} }{2} \left( \dot{R} - \zeta \frac{dU}{dR} \right)^{2}  + \frac{ c^{2} }{ 2 R^{2} }.
\end{eqnarray}
Following the procedure of classical dynamics (see \cite{LL}),
the equation of motion is derived by a {\it Routhian} given by
$ {\mathcal R} = c \dot{\theta} - L $.
It is rather complicated to deal with this equation, but
one can ignore the term including $ c $ if $ c $ can be chosen as small.
Then the equation of motion is written in a form of the so called ``instanton'' equation
\begin{eqnarray}
  \label{eq:instanton}
  \dot{R} = \zeta \frac{dU}{dR} = 2 \zeta R.
\end{eqnarray}
By using this, $ K_{sc} $ is simply written as
\begin{eqnarray}
  \label{eq:pathinstanton}
  K_{sc} = \exp \left[ - \frac{ c^{2} }{ 2 h } \int_{t_{i}}^t \frac{dt}{R^2} \right].
\end{eqnarray}
Furthermore, noting $ \frac{dt}{dR} =  \frac{1}{2 \zeta R} $,
it follows that the transition probability from the initial point $ R_{i} $ to the final one $ R_{f} $ is
\begin{eqnarray}
  \label{eq:finalpath}
  K_{sc} ( R_{f}, t \vert R_{i}, t_{i} ) = 
  \exp \left[ - \frac{c^2}{ 4 \zeta h}\int_{R_i}^{R_{f}} \frac{dR}{R^3} \right] \nonumber \\
  = \exp \left[ - \frac{c^{2}}{8 \zeta h} \left( \frac{ 1 }{ R_{i}^{2} } - \frac{ 1 }{ R_{f}^{2} } \right) \right].
\end{eqnarray}

As the final step, attaching the index of mode for the orbit,
we write $ R $ as $ R ( k ) $,
and taking the product over all modes $ k $, we obtain 
\begin{eqnarray}
  K_{sc} ( \{ R_{f} \} , t \vert \{ R_{i} \} , t_{i} ) = \prod_{k}
  \exp \left[ - \frac{ c_{k}^{2} }{ 4 {\zeta}_{k} h } \int_{R(k)_i}^{R(k)_{f}} \frac{dR(k)}{R(k)^3} \right] \nonumber \\
  = \prod_{k} \exp \left[ - \frac{ c_{k}^{2} }{ 8 {\zeta}_{k} h } \left( \frac{ 1 }{ R(k)_{i}^{2} } - \frac{ 1 }{ R(k)_{f}^{2} } \right) \right],
\end{eqnarray}
where $ c_{k} $, $ {\zeta}_{k} $ indicate that the quantities correspond to mode $ k $.

\section{Summary}

We studied a stochastic theory for the generalized Schr{\"o}dinger equation by using a method of eigenfunction expansion.
The present approach would have an advantage in that once one starts with the expansion for the wave function (order parameter) 
in terms of the set $ \{a_n\} $, one can always obtain the Langevin equation in a very general way.
Thus this approach would provide an efficient way to investigate a wide class of systems that can be described by the generalized Schr{\"o}dinger equation.

By using the functional (or path) integral formalism, the Langevin equation results in the FP equation
based on the assumption of Gaussian white noise for the fluctuation.
We paid particular attention to the calculation of the functional Jacobian,
which is simply incorporated in the action functional. 
As a consequence of this procedure, we arrived at the specific form of the FP equation (\ref{eq:FP}),
which is relevant to our subsequent analyses of the distribution function.

The analyses of the distribution function have been given for the case of the unperturbed Hamiltonian,
which can be treated within two categories:
(i) by using the expansion with respect to the dissipation constant $ {\alpha} $,
and (ii) by adopting the asymptotic limit of ``zero'' diffusion ($ h \sim 0 $).
Our analyses have an advantage in that we can obtain a concise analytic form
for the distribution function and the transition probability. 
In this connection, the FP equation that was developed in \cite{Someda} seems to be
a complicated way to obtain a simple form of the distribution function.
As a potentially useful application of our method, we mention the problem of a calculation for
various sorts of transport coefficients in nonequilibrium statistical physics that is inspired by the generalized 
Schr{\"o}dinger equation \cite{Imry}.  

An immediate application of the present formalism could be the calculation of conductivity for the 
superconducting fluctuation current just above the critical temperature.
The FP equation (\ref{eq:FP}) is expected to play a key role in calculating
the time-dependent average for the current leading to the conductivity.
The details of this topic will be discussed in a forthcoming paper.

\begin{appendix}

\section{The calculation of the functional Jacobian}
\label{append:functionaljacobi}

To evaluate $ \det \left[ \frac{{\delta}F_{n}}{{\delta}a_{n}} \right] $,
we start with a general procedure that was developed in \cite{Dashen}.
Let us write $ F = {\gamma} \frac{da}{dt} - A $,
and define $ F = {\gamma} \frac{db}{dt} $.
We note the relation
\begin{eqnarray}
  \label{eq:appenddeter}
  \det \left[ \frac{ {\delta} F}{ {\delta} a} \right] = \det \left[ \frac{ {\delta} \dot{b} }{ {\delta} b} \frac{ {\delta} b}{ {\delta} a} \right]
  = \left( \det \left[ \frac{ {\delta} \dot{b} }{ {\delta} b} \right] \right) \left( \det \left[ \frac{ {\delta} b }{ {\delta} a} \right] \right).
\end{eqnarray}
The first factor $ \det \left[ \frac{ {\delta} \dot{b} }{ {\delta} b} \right] = \det \left[ \frac{d}{dt} \right] = C $
can be omitted because it is simply a divergent factor.
Then we have the integral equation
\begin{eqnarray}
  \label{eq:appendintegral}
  {\gamma} b(t) =  {\gamma} a(t) - \int_{0}^{t} A \left[ a( {\tau} ) \right] d {\tau},
\end{eqnarray}
which is rewritten by using the step function,
\begin{eqnarray}
  \label{eq:appendintegraltheta}
  {\gamma} b(t) =  {\gamma} a(t) - \int_{0}^{T} \theta( t - {\tau} ) A \left[ a( {\tau} ) \right] d {\tau}.
\end{eqnarray}
Using this form, the interval of integration can be converted to $ [ 0, T ] $,
which coincides with the interval in the path integral (\ref{eq:resultingpath}).
Then, Eq. (\ref{eq:appendintegraltheta}) is expressed as a discrete form:
\begin{eqnarray}
  \label{eq:appendintegraldiscre}
  {\gamma} b( t_{i} ) =  {\gamma} a( t_{i} ) - \sum_{j} \theta( t_{i} - {\tau}_{j} ) A \left[ a( {\tau}_{j} ) \right] \varepsilon.
\end{eqnarray}
By carrying out the differential with respect to $ a ( t_{k} ) $, we obtain
\begin{eqnarray}
  \label{eq:appenddiffeo}
  {\gamma} \frac{ {\partial} b( t_{i} ) }{ {\partial} a( t_{k} ) } 
  =  {\gamma} \frac{ {\partial} a( t_{i} )}{ {\partial} a( t_{k} ) } - \sum_{j} \theta( t_{i} - {\tau}_{j} ) \frac{ {\partial} A \left[ a( {\tau}_{j} ) \right] }{ {\partial} a( t_{k} ) } \varepsilon,
\end{eqnarray}
which is reduced to
\begin{eqnarray}
 \frac{ {\partial} b( t_{i} ) }{ {\partial} a( t_{k} ) } = {\delta}_{ik} - \frac{1}{\gamma}\theta( t_{i} - t_{k} ) \frac{ {\partial} A \left[ a( t_{k} ) \right] }{ {\partial} a( t_{k} ) } \varepsilon.
\end{eqnarray}
The matrix represented by these elements is a triangular matrix.
For $ t_{i} > t_{k} $, it follows that
\begin{eqnarray}
  \label{eq:appendtheta1}
  \frac{ {\partial} b( t_{i} ) }{ {\partial} a( t_{k} ) } = - \frac{1}{\gamma} \frac{ {\partial} A \left[ a( t_{i} ) \right] }{ {\partial} a( t_{i} ) } \varepsilon.
\end{eqnarray}
For $ t_{i} = t_{k} $, noting $ \theta( t_{i} - t_{k} ) = 1 / 2 $, which is a Dirichlet discontinuous factor \cite{Iwanami}, we get
\begin{eqnarray}
  \label{eq:appendtheta12}
  \frac{ {\partial} b( t_{i} ) }{ {\partial} a( t_{i} ) } = 1 - \frac{1}{2 \gamma} \frac{ {\partial} A \left[ a( t_{i} ) \right] }{ {\partial} a( t_{i} ) } \varepsilon.
\end{eqnarray}
Here it is crucial to have the factor $ \frac{1}{2} $.
The determinant, except for the infinite factor $ C $, is given as 
\begin{eqnarray}
  \label{eq:appendeter}
  \det \left[ \frac{ {\delta} F }{ {\delta} a }  \right] &=& \det \left[ \frac{ {\delta} b }{ {\delta} a }  \right]  \nonumber \\ 
  &=& \prod_{i} \left( 1 - \frac{1}{2{\gamma}} \frac{ {\partial} A[ a(t_{i}) ] }{ {\partial} a( t_{i} ) }  \varepsilon \right),
\end{eqnarray}
which turns out to be 
\begin{eqnarray}
  \label{eq:appenddeter2}
  \det \left[ \frac{ {\delta} b }{ {\delta} a } \right] = \exp \left[ - \frac{1}{2 {\gamma}} \int_{0}^{T} \frac{{\partial}A[a(t)]}{{\partial}a(t)} dt \right].
\end{eqnarray}
Therefore, by substituting $ A = \frac{\partial E}{\partial a_n^{*}} $, we have the functional Jacobian,
\begin{eqnarray}
  \label{eq:determinant1append}
  {\rm{det}} \left[ \frac{{\delta}F_{n}}{{\delta}a_{n}} \right] = {\rm{exp}} \left[ - \frac{1}{ 8 \gamma} \int_{0}^{T} \left( \frac{ {\partial}^{2} E }{ {\partial} X_{n}^{2} } + \frac{ {\partial}^{2} E }{ {\partial} Y_{n}^{2} } \right) dt \right].
\end{eqnarray}
Also we have the complex conjugate of (\ref{eq:determinant1append}),
\begin{eqnarray}
  \label{eq:determinant2append}
  {\rm{det}} \left[ \frac{ {\delta} F_{n}^{*} }{ {\delta} a_{n}^{*} } \right] = {\rm{exp}} \left[ - \frac{1}{ 8 {\gamma}^{*} } \int_{0}^{T} \left( \frac{ {\partial}^{2} E }{ {\partial} X_{n}^{2} } + \frac{ {\partial}^{2} E }{ {\partial} Y_{n}^{2} } \right) dt \right].
\end{eqnarray}
By summing up (\ref{eq:determinant1append}) and (\ref{eq:determinant2append}), we obtain the terms
that are proportional to $ M $ in the text.

\end{appendix}

\end{document}